\documentclass[sigconf,9pt]{acmart}

\renewcommand\footnotetextcopyrightpermission[1]{}
\setcopyright{none}
\acmDOI{}
\acmISBN{}
\acmConference[]{}
\acmYear{2026}
\copyrightyear{}
\acmPrice{}

\usepackage{algorithm}
\usepackage{algpseudocode}
\usepackage{graphicx}
\usepackage{textcomp}
\usepackage{xcolor}
\def\BibTeX{{\rm B\kern-.05em{\sc i\kern-.025em b}\kern-.08em
    T\kern-.1667em\lower.7ex\hbox{E}\kern-.125emX}}
\usepackage{subcaption}
\usepackage{xspace}
\usepackage{tikz}

\newcommand{\FBF}[1]{\noindent\textbf{#1} }
\newcommand{\CIRCLEDB}[1]{%
  \tikz[baseline=(char.base)]{
    \node[shape=circle, draw, fill=gray!120, text=white, inner sep=0.5pt] (char) {#1};}}

\newcommand{\name}{\textsf{Q-StaR}\xspace}
\newcommand{\namea}{\textsf{N-Rank}\xspace}
\newcommand{\nameb}{\textsf{BiDOR}\xspace}

\begin{document}
\title{
  \name: A Quasi-Static Routing Scheme for NoCs
}

\author{Yang Zhang}
\affiliation{%
  \institution{Tsinghua University}
  \city{Beijing}
  \country{China}
}
\email{zhang-y22@mails.tsinghua.edu.cn}

\author{Yiren Zhao}
\affiliation{%
  \institution{University of Toronto}
  \city{Toronto}
  \country{Canada}
}
\email{yiren.zhao@mail.utoronto.ca}

\author{Xu Wang}
\affiliation{%
  \institution{Lanzhou University}
  \city{Lanzhou}
  \country{China}
}
\email{xw2025@lzu.edu.cn}

\author{Fengyuan Ren}
\affiliation{%
  \institution{Tsinghua University}
  \city{Beijing}
  \country{China}
}
\email{renfy@tsinghua.edu.cn}

\begin{abstract}
In networks-on-chip, static routing schemes are favored for their simplicity and predictability, but they cannot effectively balance network load due to the unawareness of runtime load distribution.
\name discovers two factors (topology and traffic distribution) that determine the long-term trend of load distribution, and proposes \namea to extract this trend.
The obtained information is used to guide \nameb's route selection at runtime, thereby improving load balancing while retaining simplicity and predictability.
Simulation validates that \name significantly outperforms the typical dimension-order routing (throughput under uniform traffic improved by 42.9\%, and mean/maximum latency under realistic workloads reduced by 86.4\%/95.3\%).
\end{abstract}

\maketitle
\pagestyle{plain}

\section{Introduction}

Modern systems-on-chip (SoCs) rely on networks-on-chip (NoCs) as their internal communication backbone. The NoC connects dozens, hundreds, or even more components through a structured network of routers and channels, and directly affects the overall performance, power efficiency, and scalability of the SoC.

Although various topologies have been proposed for NoCs, the two-dimensional mesh (2DMesh) is still the dominant choice in production chips for its simplicity and ease of 2D layout \cite{arm2021neoversen2,nvidia2019a011,Cerebraswse3}. On top of this topology, routing algorithms are expected to be both simple and provably safe. Consequently, dimension-order routing (DOR, known as XY or YX routing in 2DMesh networks), is widely adopted for its favorable properties like low hardware cost, in-order transmission, and guaranteed deadlock freedom \cite{Dally2004principles}.
Despite its simplicity and predictability, DOR statically makes routing decisions unaware of load distribution and may incur severe load imbalance at runtime, thus delivering poor worst-case (or even average-case) performances.

To better balance load through routing, previous research has explored two main directions. Oblivious routing schemes \cite{o1turn,nesson1995romm,valiant1981universal} randomly disperse traffic among different paths, independent of instantaneous load.
They introduce moderate hardware overhead and have acceptable complexity, but the random path selection either provides limited load balancing or introduces unnecessary detours, resulting in poor performance in certain cases.
Adaptive routing schemes \cite{ma2011dbar,oddeven,turnmodel}, in contrast, monitor network conditions at runtime and steer packets toward light-loaded regions.
They can effectively balance network load among alternative paths, improving the overall performance under heavy or skewed traffic.
However, the costly information collection, complex control logic, and additional states make adaptive routing hard to deploy on-chip.

In NoC routing design, there exists a long-standing dilemma.
Routing schemes are expected to be static (like DOR) for simplicity and predictability, while at the same time, they are expected to be dynamic (like adaptive routing) for better load balancing at runtime.
Motivated by this, this paper attempts to find a middle ground: \textit{enhance NoC routing with long-term trends of load distribution}, so that it can retain the strengths of static routing algorithms while balancing network load more effectively at runtime.

In this work, we propose \name, a quasi-dynamic routing scheme that bridges the gap between purely static and fully dynamic routing.
\name is built on the observation that the load distribution in a NoC is not entirely unpredictable.
Instead, it is strongly influenced by two factors: the topology, which determines how traffic converges and diverges across different nodes, and the traffic distribution, which determines how traffic load is distributed within the network.
Luckily, neither of these factors is hard to obtain in NoCs: the topology is predetermined, while the traffic distribution can be inferred from the communication pattern of upstream workloads.
Compared with dynamically collected load information, these factors capture the long-term trends of load distribution.
Encoding them into a compact metric, we can bias routing decisions proactively toward less-loaded regions, without relying on runtime detection or complex control loops.

Based on these observations and insights, \namea is proposed to extract information from the topology and traffic distribution.
It uses an evolutionary model to imitate the runtime lifespan of traffic on the topology, and nodes that are more likely to experience heavy load during the evolution will be assigned higher $w_{NR}$s.
Guided by \namea, \nameb\ is developed.
For each source-destination pair, it selects between two dimension-order routes (XY and YX) and chooses the one with the lower cumulative $w_{NR}$.
This enables it to avoid heavily loaded nodes as much as possible, thereby balancing load better at runtime, and as $w_{NR}$s only captures long-term trends, the routing decisions stay static most of the time.

Together, \namea and \nameb\ form \name.
It makes swift and deterministic routing decisions at runtime, retaining the strengths of simplicity and predictability of static routing algorithms.
Simultaneously, the routing decisions are guided by long-term load information, improving the load-balancing capabilities.

\begin{figure*}
    \begin{subfigure}[b]{0.18\linewidth}
        \vspace{0pt}
        \centering
        \includegraphics[page=6,width=0.9\linewidth]{figures/idea_design.pdf}
        \caption{Load distribution.}
        \label{subfig:2m_un_layout}
    \end{subfigure}
    \hfill
    \begin{subfigure}[b]{0.26\linewidth}
        \vspace{0pt}
        \centering
        \includegraphics[width=\linewidth]{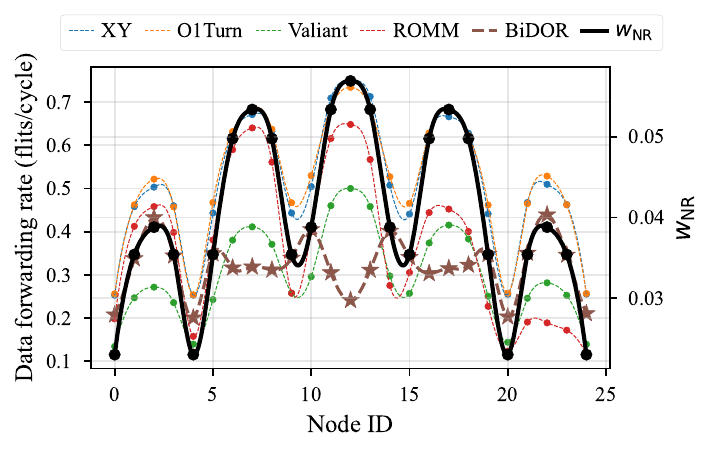}
        \caption{2DMesh + Uniform.}
        \label{subfig:2m_un}
    \end{subfigure}
    \begin{subfigure}[b]{0.26\linewidth}
        \vspace{0pt}
        \centering
        \includegraphics[width=\linewidth]{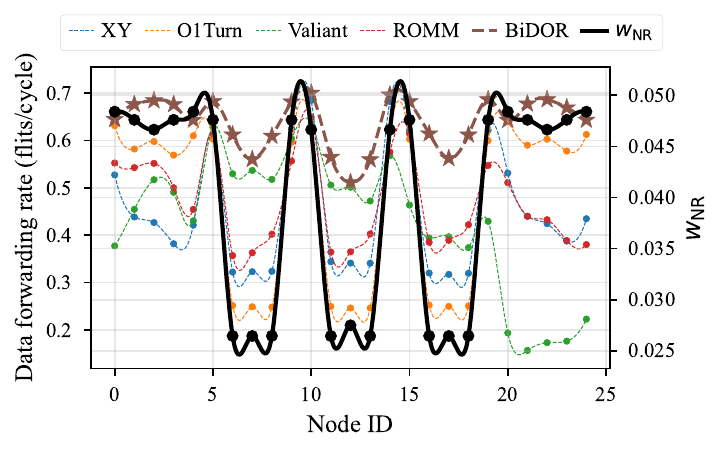}
        \caption{2DMesh (edge I/O) + Uniform.}
        \label{subfig:eb_un}
    \end{subfigure}
    \begin{subfigure}[b]{0.26\linewidth}
        \vspace{0pt}
        \centering
        \includegraphics[width=\linewidth]{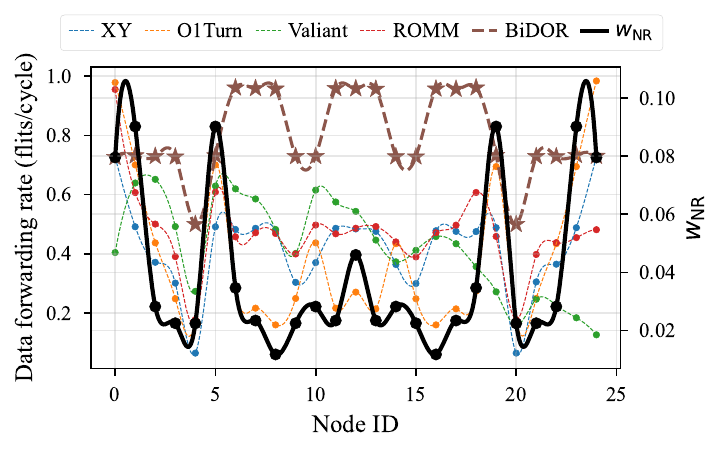}
        \caption{2DMesh (edge I/O) + Overturn.}
        \label{subfig:eb_ov}
    \end{subfigure}
    \caption{
    (a) load distribution on different nodes (2DMesh + Uniform + XY routing);
    (b), (c), (d) load distribution in different topologies, with different traffic patterns.
    }
    \label{fig:load_distri}
\end{figure*}

Experiments validate that \namea can precisely anticipate the trend of runtime load distribution, and guided by \namea, \nameb\ can balance load effectively.
It's found that \name achieves performance similar to that of adaptive routing schemes, significantly outperforming typical deterministic routing (DOR).
Under uniform traffic, it achieves 42.9\% higher throughput than DOR, and under realistic workloads, the mean and maximum latencies are reduced by 86.4\% and 95.3\%, respectively.

\section{Background and Motivation} \label{sec:moti}

\subsection{Routing Algorithms in NoCs} \label{subsec:bg_routing}

\FBF{Deterministic routing.}In deterministic routing, the path a packet takes is completely determined by predefined rules, usually based on the coordinates or locations of source and destination ports. The selected path remains fixed for each source-destination pair $\langle s, d \rangle$, regardless of runtime conditions. 
A well-known deterministic routing scheme is Dimension-Order Routing (DOR), also known as XY (or YX) routing in 2DMesh networks \cite{dor}. DOR delivers packets by traversing network dimensions in a predetermined sequence. For example, in XY routing, a flit\footnote{A flit (flow control unit) is the smallest data unit in a NoC. A data packet consists of multiple flits, and the header flit determines the path for all subsequent flits in the packet.} first traverses along the X-dimension until it reaches the destination column, and then proceeds along the Y-dimension (YX routing follows the reverse order).

Besides extreme simplicity and negligible hardware overhead, deterministic routing also offers other desirable features, such as in-order transmission, simple deadlock-avoidance, etc. However, these advantages come at the cost of poor load balancing, leading to poor performance in adverse or worst-case scenarios.

\FBF{Oblivious routing.}Oblivious routing randomly disperses traffic from $s$ to $d$ among multiple paths.
Here, we take several oblivious routing algorithms in 2DMesh networks as examples.
In O1Turn, such paths include the XY and YX routes \cite{o1turn}.
ROMM randomly selects one intermediate node within the \textit{minimum rectangle}\footnote{The rectangle enclosed by the XY route and YX route, all minimum paths are encircled within this rectangle.}, flits are first routed to this node, then towards the destination \cite{nesson1995romm}.
Similarly, Valiant randomly selects an intermediate node within the whole network, which can better distribute traffic load in the worst case, but sacrifices performance in benign cases \cite{valiant1981universal}.

By dispersing traffic load randomly, oblivious routing can achieve load-balancing to some extent.
However, without a clear knowledge of load distribution, such dispersion is oblivious and either has a limited effect or sacrifices performance in benign cases.

\FBF{Adaptive routing.}Extensive research about adaptive routing has been conducted \cite{singh2005load,gratz2008regional,ma2011dbar,cbcm}.
By adjusting routing decisions according to runtime load information, adaptive routing aims to balance load better and improve network performance.
However, in practice, the collected information may be insufficient, imprecise, and outdated, which diminishes the final effect of adaptive routing schemes \cite{jiang2009indirect}.
Moreover, the complicated and dynamic path selection makes deadlock a thorny issue, which has to be addressed by dedicated mechanisms such as turn models \cite{turnmodel,oddeven,fu2011abacus} and VC-based mechanisms \cite{dally1993deadlock,duato1993new}.
Besides, the out-of-order transmission inherent in oblivious and adaptive routing may incur additional costs of packet re-ordering, which is also non-negligible in on-chip networks  \cite{gomez2007deterministic}.

\FBF{Static, dynamic, or something else?}At this point, the design of routing algorithms appears to face a dilemma between two main categories: static (deterministic) and dynamic (oblivious or adaptive) schemes.
Static routing offers simplicity, low latency, and predictable behavior, whereas dynamic routing enhances flexibility by adjusting decisions according to runtime load distribution. 

However, these two approaches may not represent the full design space. Even dynamic routing algorithms cannot make optimal decisions when load information is insufficient and potentially inaccurate. Thus, it may be acceptable to settle for second best: \textbf{using quasi-dynamically anticipated (rather than dynamically observed) load information to guide the routing decisions}.
As we will demonstrate in \S \ref{subsec:motiexp}, this approach is both effective and feasible in NoCs.

\subsection{Anticipating Load Distribution in NoCs} \label{subsec:motiexp}

\FBF{Key factors that set the tone of load distribution.}To better understand the runtime load distribution in NoCs, a simple experiment is conducted (the detailed setup is described in \S \ref{subsec:setup}).
We first investigate the load distribution in a NoC with a 5x5 2DMesh topology under uniformly distributed traffic.
Figure \ref{subfig:2m_un_layout} shows the load distribution on different nodes (measured by the node's total data forwarding rate), with XY routing adopted.
It can be observed that nodes at central positions are more likely to withstand higher loads, while corner nodes typically have lower loads.
As we adopt other oblivious routing algorithms (Figure \ref{subfig:2m_un}), similar trends can be observed: central nodes (7,11,12,13,17) are always heavily loaded, while edge ones (0,4,20,24) are always lightly loaded.
When we adopt a different topology (2DMesh with only edge nodes providing I/O ports\footnote{This configuration is common in I/O-intensive SoCs like switching chips, in which I/O modules are typically arranged around the edges of the NoC.}) or traffic pattern (overturn), the load distribution are different (Figure \ref{subfig:eb_un} and \ref{subfig:eb_ov}), but similar trends across different routing algorithms can still be observed.

From the results, one common point can be noticed: \textit{in a NoC with a specific \textbf{topology} and \textbf{traffic distribution}, the load distribution shows similar trends with different routing algorithms adopted}.
This aligns well with intuition.
In a 2DMesh network, flits are more likely to pass through the central nodes than the corner nodes, and if the traffic is uniformly distributed, central nodes are more likely to become heavily loaded.
Similarly, with a different topology or traffic distribution, the load distribution will also exhibit a certain trend.

\begin{figure}
    \begin{subfigure}[t]{0.49\linewidth}
        \vspace{0pt}
        \centering
        \includegraphics[width=\linewidth]{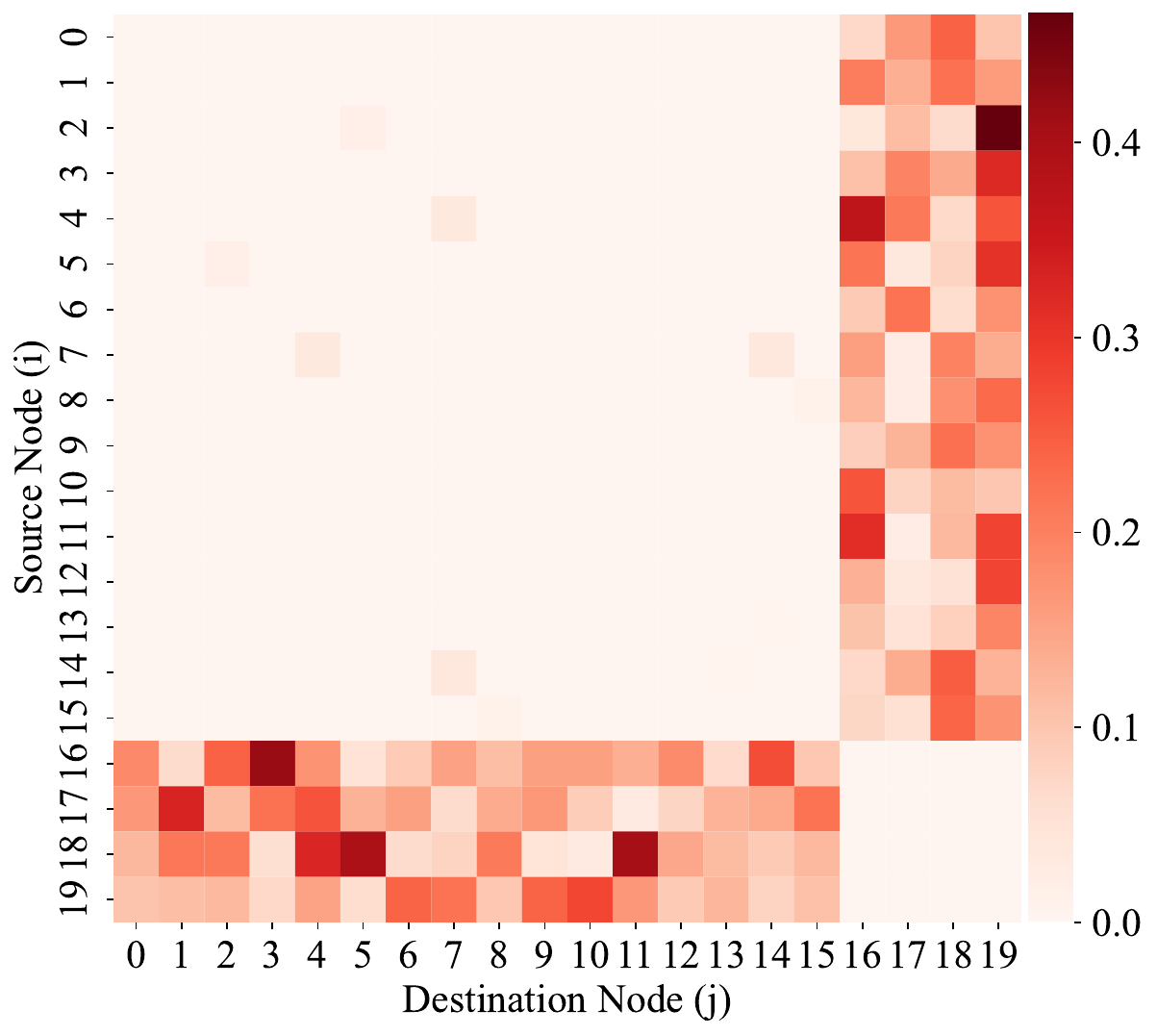}
        \caption{The switching chip of a leaf switch in a Clos network.}
        \label{subfig:switchpat}
    \end{subfigure}
    \begin{subfigure}[t]{0.49\linewidth}
        \vspace{0pt}
        \centering
        \includegraphics[width=\linewidth]{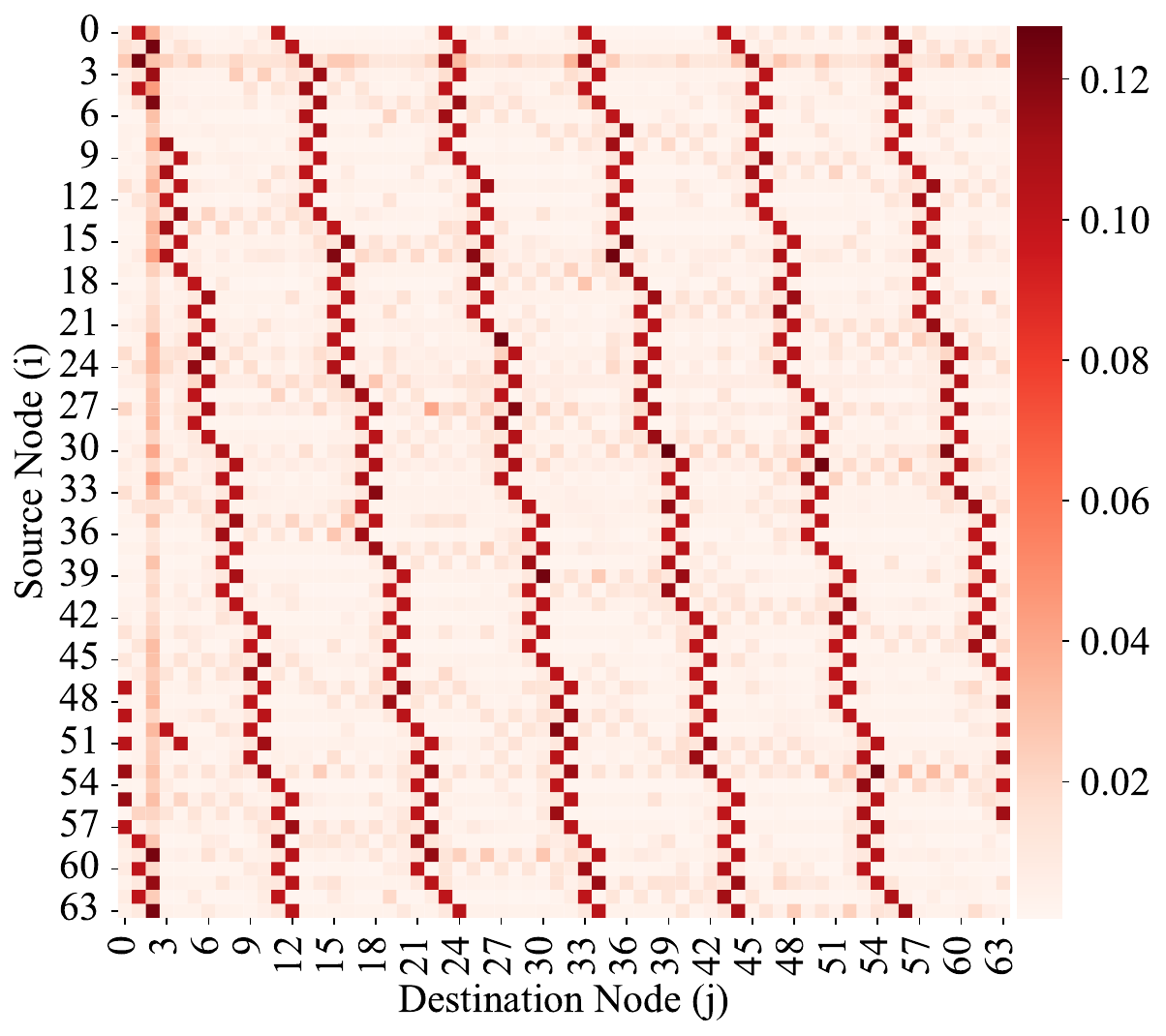}
        \caption{A 64-core system executing the Blackscholes task.}
        \label{subfig:blackscholes}
    \end{subfigure}
    \caption{Heatmap of traffic from source nodes (Y-axis) to destination nodes (X-axis), in different scenarios.}
    \label{fig:trafficdistri}
\end{figure}

\FBF{Obtaining the topology and traffic distribution in NoCs.}In NoCs, the topology is pre-determined at design time and can be easily obtained.
Network traffic is not known in advance, but the pattern it exhibits depends on the communication pattern of the upstream workload.
For a switching chip, as it is installed in upper-layer systems (data centers, HPC clusters, etc.), the traffic in it is determined by the communication requirements of the upstream system.
For a computational chip, as each NoC node serves one or several computational cores, the traffic distribution is determined by the task parallelism and communication pattern among cores.
Figure \ref{fig:trafficdistri} shows our statistics about the traffic distribution in different SoCs.
It can be found that whether in a switching chip (\ref{subfig:switchpat}) or a computational chip (\ref{subfig:blackscholes}), the traffic may show significant patterns.
With some knowledge of the workload's communication pattern, it's possible to predict the traffic distribution accurately.

In SoCs, it's possible to obtain the workload's communication pattern without much effort.
In a multi-core system, the communication pattern can be determined at compile time.
When a task is activated, the corresponding information can be leveraged to predict the subsequent traffic distribution.
In a switching chip, once the switch is installed in the upper-layer network, the traffic distribution is determined and can be obtained through analysis or statistics.

\FBF{Opportunities arising from quasi-dynamic information.}Compared with dynamically collected load information at runtime, the topology and traffic distribution information are \textit{\textbf{quasi-dynamic}: they remain unchanged, or change only at long time scales}.
This property allows mechanisms to carefully collect and process information offline, without affecting the performance of normal data transmission.
Besides, the acquired knowledge can be hard-coded into routing algorithms, ensuring swift and deterministic routing decisions at runtime.
\section{Design} \label{sec:design}
\begin{figure}
    \centering
    \includegraphics[page=1,width=0.9\linewidth]{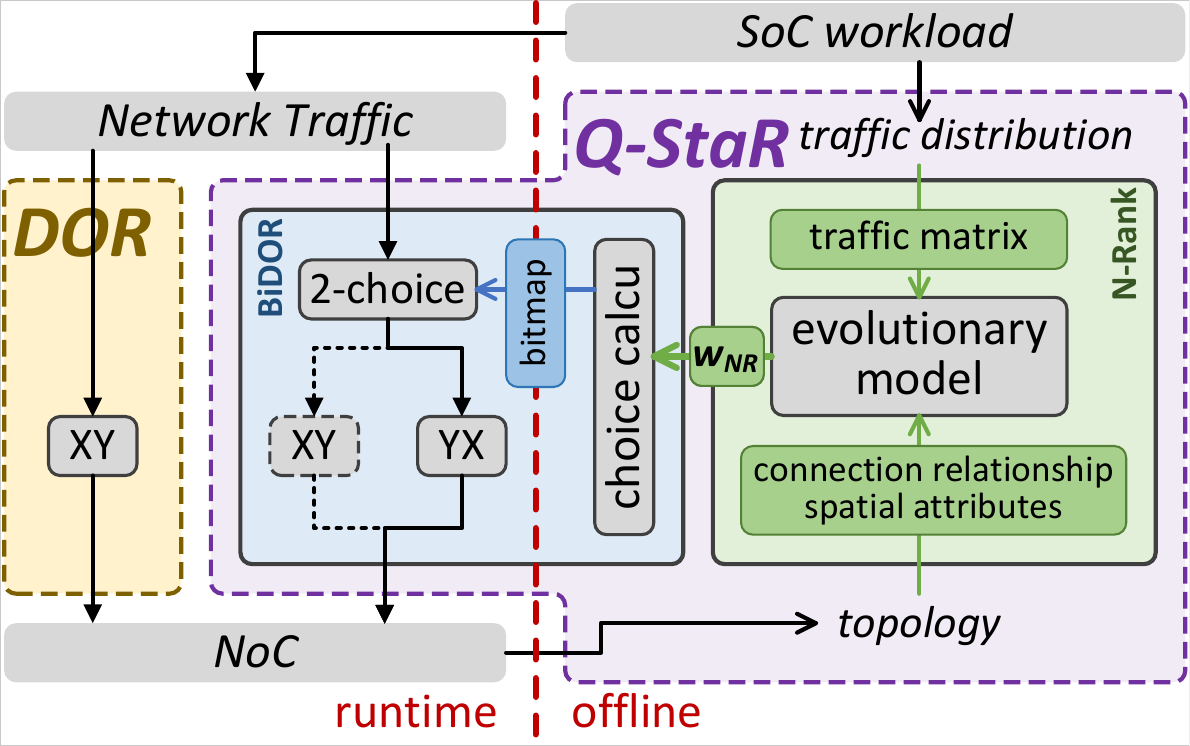}
    \caption{Workflow of \name (compared with DOR).}
    \label{fig:overview}
\end{figure}
\subsection{Overview}

Inspired by above observations and insights, we develop \name.
As shown in Figure \ref{fig:overview}, it consists of two components:

\textit{\namea} uses an evolutionary model to extract load information from the topology and traffic distribution.
It determines the $w_{NR}$ of each node, which reflects the node's likeliness of being heavily loaded at runtime.
Since traffic distribution changes infrequently, there is ample time for \namea to complete this task offline meticulously.

\textit{\nameb} chooses between two DOR choices according to the information provided by \namea.
As $w_{NR}$s are only updated offline, the choice can be calculated offline and hard-coded as bitmaps for quick runtime lookup.

\namea and \nameb\ together constitute \textit{\name}.
It has similar runtime behavior to DOR, but is also guided by quasi-dynamic load information.
This allows it to better balance the load at runtime while retaining the advantages of static routing.
In the following subsections, detailed designs of \namea (\S \ref{sec:noderank}) and \nameb\ (\S \ref{sec:flipdor}) are presented.

\subsection{\namea} \label{sec:noderank}

\namea considers two factors: topology and traffic distribution.

From the topology, the \textit{connection relationships} among nodes are obtained, and the topology's \textit{spatial attributes} are also implicitly leveraged.
The connection relationship is formalized as each node $n$'s upstream (downstream) set $U^n$ ($D^n$), i.e., the set of nodes that have a channel towards (from) $n$.
From the traffic distribution, we focus on the spatial distribution of traffic.
This distribution is formalized as a \textit{traffic matrix} $T$\cite{Dally2004principles}, where each matrix element $T_{s,d}$ gives the fraction of traffic sent from node $s$ destined to node $d$.

These factors will be fed into the evolutionary model to extract the trends in runtime load information.

\subsubsection{Evolutionary Model}

\begin{figure}
        \vspace{0pt}
        \centering
        \includegraphics[page=2,width=0.85\linewidth]{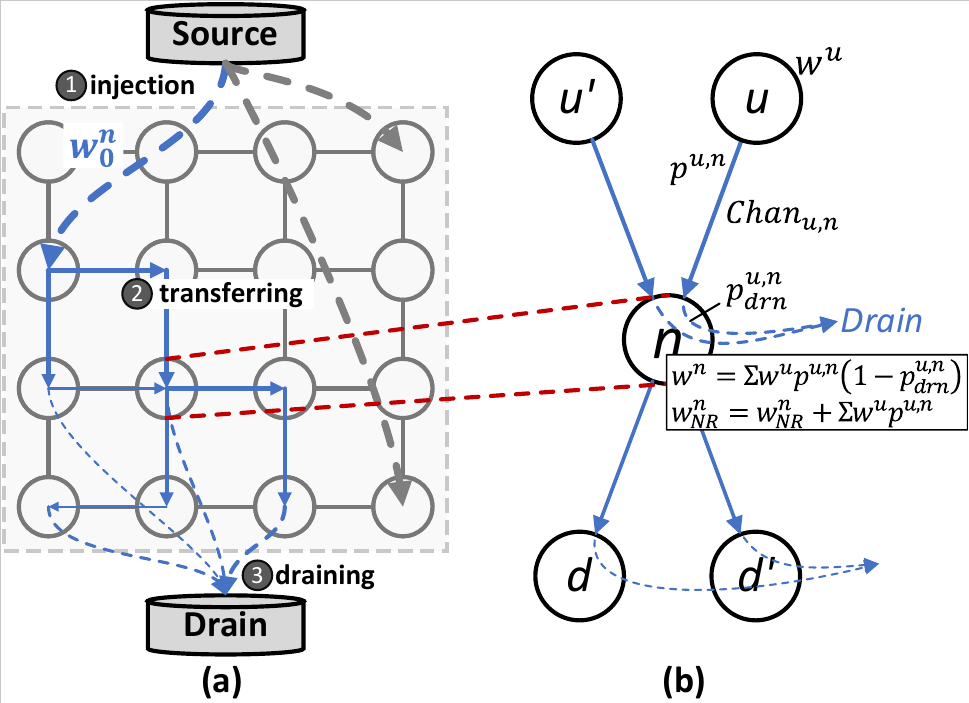}
        \caption{Evolutionary Model of \namea.}
        \label{fig:noderank}
\end{figure}

As shown in Figure \ref{fig:noderank}a, an evolutionary model is adopted, which imitates the lifespan of network traffic at runtime:
Initially, every node $n$ is \CIRCLEDB{1}\textit{injected} with a certain quantity of traffic.
After that, traffic \CIRCLEDB{2}\textit{transfers} randomly among nodes, and finally gets \CIRCLEDB{3}\textit{drained}.
When most of the traffic is drained from the network, the evolution terminates, and the accumulated load that one node $n$ has experienced during the evolution is quantified as its \textit{NR-weight} $w_{NR}^n$, which directly indicates $n$'s likelihood of being heavily loaded at runtime.
Based on this model, \namea works iteratively, with the following three key stages:

\FBF{Initialization.}Every node $n$ is first allocated with an initial weight $w_0^n$, and the NR-weight is also initialized with the same value.
The value of $w_0^n$ equals the proportion of traffic that is sourced from this node:
\begin{align}
    w_0^n=\sum\nolimits_{n' \in N}{T_{n,n'}}
    \label{eq:initialrank}
\end{align}

\FBF{Iteration.}Every iteration, one node transfers weight to its downstream nodes and gains new weight transferred from its upstream nodes.
Simultaneously, NR-weight records the cumulative weight it experienced.
From the perspective of node $n$, its weight $w^n$ and NR-weight $w_{NR}^n$ are updated as follows:
\begin{align}
    &w^n        \gets \sum\nolimits_{u \in U^n}{w^up^{u,n}(1-p_{drn}^{u,n})} \\
    &w_{NR}^n  \gets w_{NR}^n + \sum\nolimits_{u \in U^n}{w^up^{u,n}}
\end{align}
In these equations, $p^{u,n}$ and $p_{drn}^{u,n}$ represent the transition probability and draining probability from node $u$ to node $n$, respectively.
In each iteration, the weight of node $u$ is transferred to all its downstream nodes, while node $n$ receives a portion of $p^{u,n}$.
Of the transferred weights, a portion of $p_{drn}^{u,n}$ is drained and will not be transferred in subsequent iterations.
The quantification of $p^{u,n}$ and $p_{drn}^{u,n}$ requires a comprehensive consideration of the traffic matrix and the topology's spatial attributes, and will be further discussed in \S \ref{subsubsec:quantify}.

\FBF{Termination.}As the total quantity of $w^n$ in the system decreases with every iteration due to draining, the evolution is expected to converge after some iterations.
\namea terminates in two cases: the sum of $w^n$ falls below a threshold $w_{th}$, or the iteration count $iter$ exceeds a threshold $iter_{th}$.
In practice, we set $w_{th}=0.01, iter_{th}=100$, and the evolution usually terminates after about 40 iterations, indicating that \namea can converge easily in normal cases.

\subsubsection{Quantifying Relevant Variables} \label{subsubsec:quantify}

\namea attempts to define the transfer and draining probabilities with only one simple assumption about routing algorithms: \textit{they never take detours}.
Based on this assumption, $p^{u,n}$ and $p_{drn}^{u,n}$ are set following one common principle: \textit{more possibilities, higher probability}.

\FBF{Possibility set and the corresponding weight.}We first define the possibility set $P^{u,n}$ and its weight $W^{u,n}$ of the transfer along the channel from $u$ to $n$  ($Chan_{u,n}$).
\begin{figure}
    \begin{minipage}[t]{0.48\linewidth}
        \vspace{0pt}
        \centering
        \includegraphics[page=3,width=\linewidth]{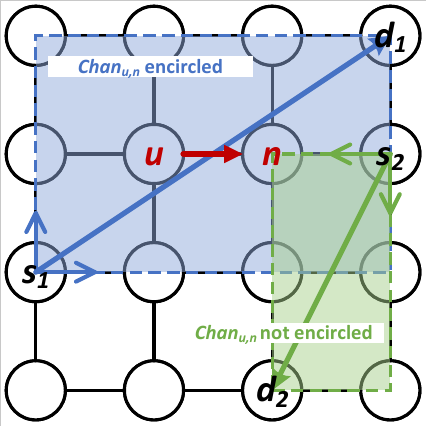}
        \caption{$\langle s_1,d_1 \rangle$'s minimum rectangle encircles $Chan_{u,n}$, thus in $P^{u,n}$, otherwise not ($\langle s_2,d_2 \rangle$).}
        \label{fig:transprob}
    \end{minipage}
    \hspace{0.02\linewidth}
    \begin{minipage}[t]{0.48\linewidth}
        \vspace{0pt}
        \centering
        \includegraphics[page=4,width=\linewidth]{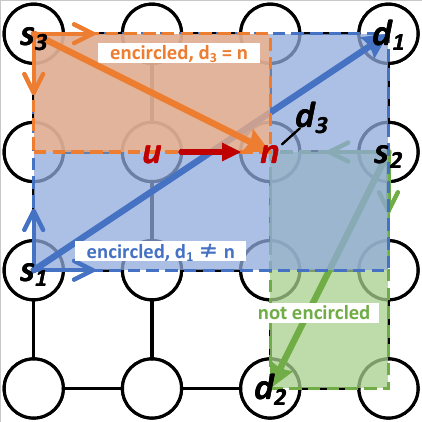}
        \caption{Besides rectangle-encircling, $\langle s,d \rangle$ in $P_{drn}^{u,n}$ should also have $n$ as destination ($\langle s_3,d_3 \rangle$).}
        \label{fig:drainprob}
    \end{minipage}
\end{figure}
The possibility set $P^{u,n}$ contains all possible node pairs $\langle s,d \rangle$ that satisfy the following constraint: the traffic from $s$ to $d$ is possible to pass through $Chan_{u,n}$ without any detouring.
Take the 2DMesh topology as an example (Figure \ref{fig:transprob}), $\langle s, d \rangle$ is in $P^{u,n}$ if and only if the minimum rectangle from $s$ to $d$ ($MinRect_{s,d}$) encircles $Chan_{u,n}$.
Further, the possibility weight $W^{u,n}$ refers to the total proportion of traffic in the possibility set, which can be obtained by summing up all the corresponding elements in the traffic matrix.
\begin{align}
    &P^{u,n} = \left\{ \langle s,d \rangle | Chan_{u,n} \subset MinRect_{s,d} \right \} \\
    &W^{u,n} = \sum\nolimits_{\langle s,d \rangle \in P^{u,n}}{T_{s,d}}
\end{align}

Similarly, the possibility set $P_{drn}^{u,n}$ and weight $W_{drn}^{u,n}$ of draining can also be determined.
Besides the constraint of encircling $Chan_{u,n}$ with the minimum rectangle, the node pair $\langle s,d \rangle$ should also have node $n$ as the destination (Figure \ref{fig:drainprob}).
\begin{align}
    &P_{drn}^{u,n} =\left\{ \langle s,d \rangle | \langle s,d \rangle \in P^{u,n}, d=n \right \} \\
    &W_{drn}^{u,n} = \sum\nolimits_{\langle s,d \rangle \in P_{drn}^{u,n}}{T_{s,d}}
\end{align}

\FBF{Transfer and draining probability.}
As the $W^{u,n}$ ($W_{drn}^{u,n}$) reflects the total proportion of traffic that is likely to select $Chan_{u,n}$ as one hop (get drained after traversing $Chan_{u,n}$), the transfer and draining probability can be calculated accordingly:
$p^{u,n}$ is the proportion of $W^{u,n}$ among all possible downstream channels, and $p^{u,n}_{drn}$ is the proportion of $W_{drn}^{u,n}$ in $W^{u,n}$.
\begin{align}
    &p^{u,n} = \frac{W^{u,n}}{\sum_{n' \in D^{u}}{W^{u,n'}}} \label{eq:pij}\\
    &p_{drn}^{u,n} = \frac{W_{drn}^{u,n}}{W^{u,n}}
\end{align}
\subsection{\nameb} \label{sec:flipdor}

\nameb\ takes the information ($w_{NR}$s) provided by \namea and makes routing decisions accordingly.
The route calculation is completed offline, and the results will be hard-coded as bitmaps for quick and deterministic runtime routing.

\subsubsection{Route Calculation}
\begin{figure}
    \centering
    \includegraphics[page=5,width=\linewidth]{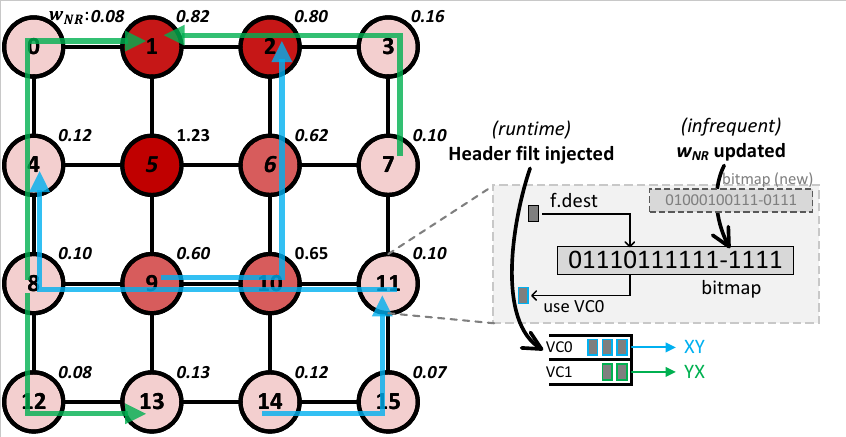}
    \caption{\nameb\ chooses between XY and YX routes, and the choice is stored as bitmaps for quick runtime lookup.}
    \label{fig:flipdor}
\end{figure}

As the left part of Figure \ref{fig:flipdor} shows, the route-selecting principle of \nameb\ is simple: \textit{compare two DOR choices (XY and YX), and find the one with smaller total $w_{NR}$.}
Using $R_0^{s,d}$ and $R_1^{s,d}$ to refer to the node sequence of the XY and YX route from $s$ to $d$, the route choice is selected as follows:
\begin{align}
    b^{s,d}=\arg\min\nolimits_{b\in\{0,1\}}\left(\sum\nolimits_{n\in R_b^{s,d}}w_{NR}^n \right)
\end{align}
If $b^{s,d}=0$, traffic from $s$ to $d$ will be routed through the XY route, otherwise ($b^{s,d}=1$), through the YX route.
Take the routing from node 11 to 4 as an example, two routes (XY:11-10-9-8-4 and YX:11-7-6-5-4) are considered.
The cost of two routes is measured by the sum of $w_{NR}$s of all nodes along the path (XY:1.57, YX:2.17).
The one with the lower cost is selected ($b^{11,4}=0$), and all traffic from node 11 to 4 follows the XY route.

Because $w_{NR}$s rarely change, the route choices can be hard-coded on each node to support swift runtime route decisions.
As shown in the right part of Figure \ref{fig:flipdor}, each node $s$ maintains a $|N|$-bitmap ($|N|$ is the total number of nodes), in which each bit represents the routing choice to the corresponding destination.
\begin{align}
    bitmap^{s}=\left[b^{s,0},b^{s,1},...b^{s,|N|-1}\right]
\end{align}
When, and only when, $w_{NR}$s are updated (when the upstream workload changes significantly), new bitmaps will be calculated and overwritten into the corresponding nodes.
In this way, \nameb\ refreshes its routing policy and `adapts' to the updated load distribution information.

\subsubsection{Runtime Routing}
When a header flit is injected at one node, its destination is used as an index to look up in the bitmap.
If the corresponding bit is set as 0 (1), the flit will be routed along the XY (YX) route afterwards.
\nameb\ uses two isolated virtual channels (VCs), each corresponding to a DOR routing choice (VC0 for XY-routed flits, and VC1 for YX-routed ones).
This not only eliminates the risk of deadlock, but also saves the need to carry the route choice in data flits.

Besides the improved performance in load-balancing (evaluated in \S \ref{sec:eval}), \nameb\ also inherits most strengths of DOR, including:
\begin{itemize}
    \item Simplicity. Compared with DOR, \nameb\ requires only one additional VC and introduces one bitmap lookup when making routing decisions. This makes it easy to deploy, and the impact on normal traffic transmission is negligible.
    \item Deadlock-free. XY and YX routing have been proven to be deadlock-free \cite{Dally2004principles}. Because traffic on XY and YX routes is isolated in separate VCs, \nameb\ is free from deadlock.
    \item In-order transmission. \nameb\ is a quasi-dynamic routing algorithm. The route between each port-pair $\langle s,d \rangle$ changes only when \namea is updated. This ensures that most traffic follows one fixed path, avoiding out-of-order transmission most of the time.
\end{itemize}
\section{Evaluation} \label{sec:eval}
\begin{figure*}
    \centering
    \includegraphics[width=0.95\linewidth]{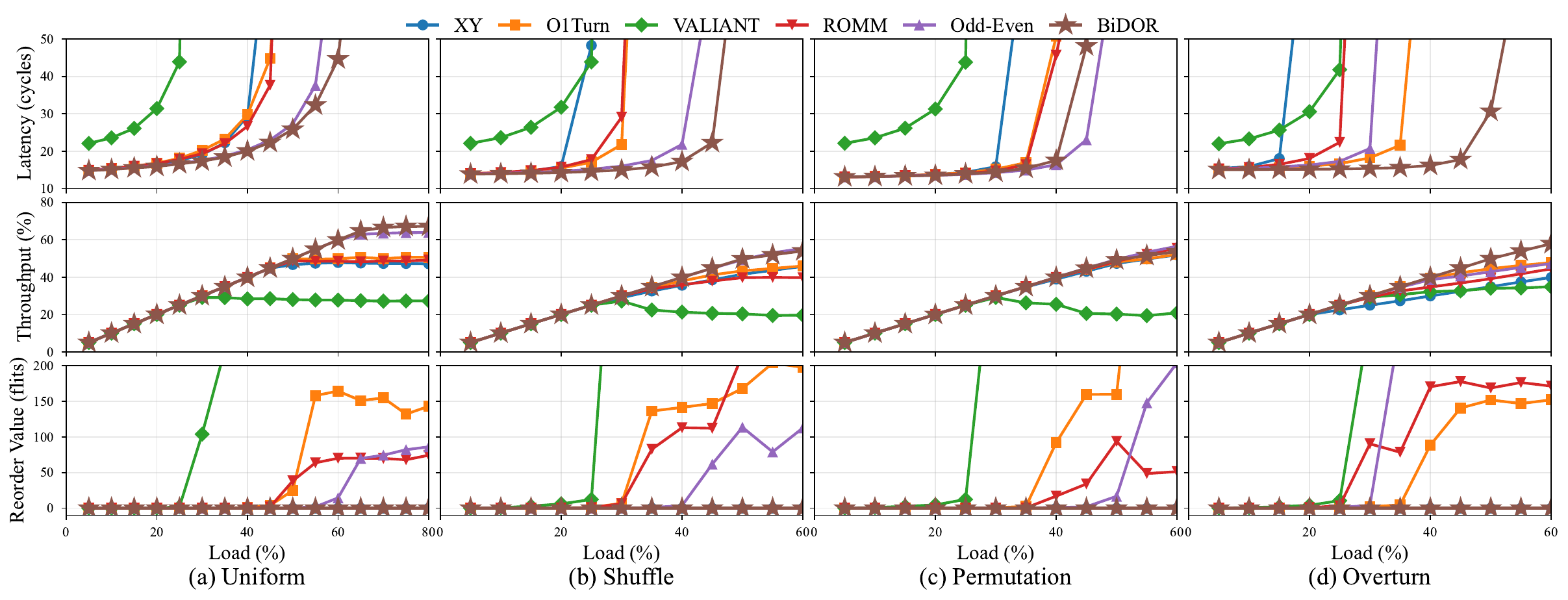}
    \caption{Comparison of different routing algorithms under typical traffic patterns.}
    \label{fig:synthetic}
    \vspace{-10pt}
\end{figure*}

\begin{figure}
    \vspace{0pt}
    \centering
    \includegraphics[width=0.9\linewidth]{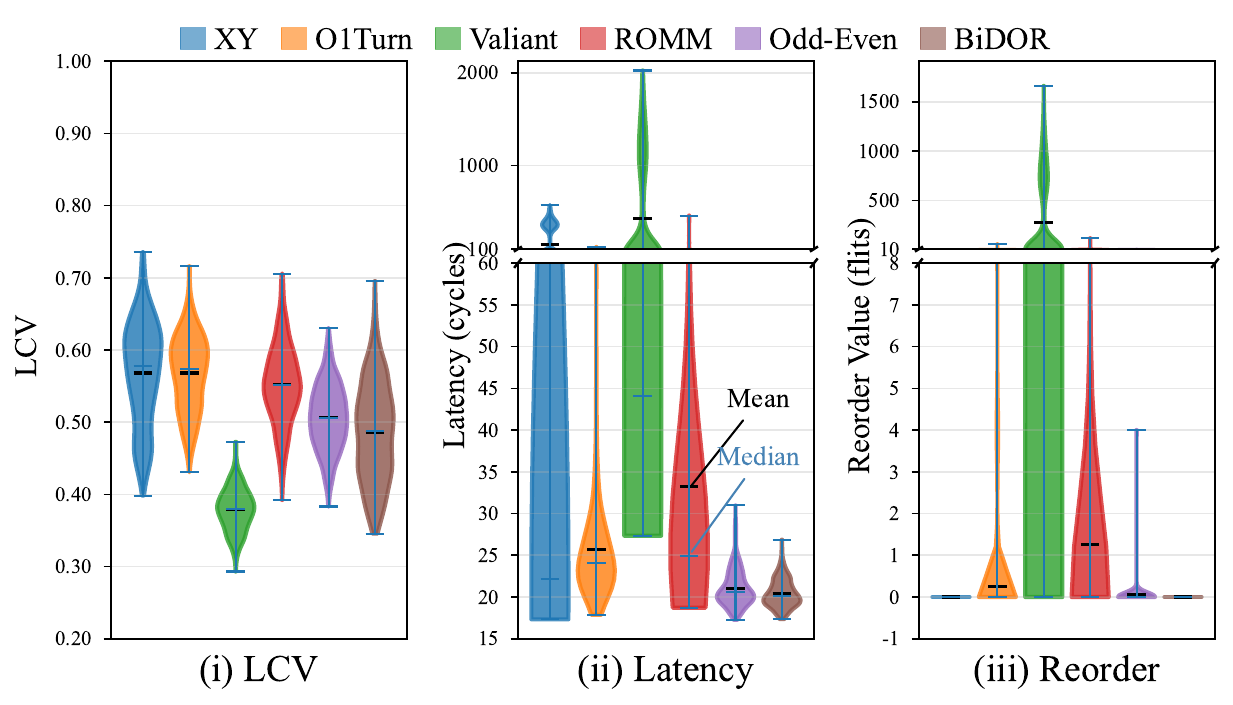}
    \caption{Different routing algorithms' performance under realistic workload.}
    \label{fig:switchtrace}
    \vspace{-10pt}
\end{figure}

To evaluate the performance of \name, simulations are conducted on BookSim2, a cycle-accurate simulator \cite{booksim2}.
The experiment is set up as explained in \S \ref{subsec:setup}.
\name's performance is compared with other routing schemes under synthetic (\S \ref{subsec:synthetic}) and realistic (\S \ref{subsec:trace}) traffic.

\subsection{Experiment Setup} \label{subsec:setup}

The NoC is built as a 5x5 2DMesh, which provides 20 I/O ports connected to its edge nodes (5 I/O ports at each edge).
The NoC routers are input-queued; each has 4 bi-directional ports, either used for I/O or interconnection with other nodes.
The buffer is allocated in the granularity of data flits (wormhole), and credit-based flow control is adopted to avoid data loss due to buffer overflow.
Each input port has a total buffer of 64 flits, shared by 2 virtual channels.
For simplicity, we assume that the routing decision can be made within one cycle and each hop has a base latency of 2 cycles (one for logic processing, one for channel traversal).\footnote{As all routing algorithms (except adaptive routing) are of similar complexity, this assumption is without loss of generality and will only introduce a favorable bias to the performance of adaptive routing.}

\name is implemented as described in \S \ref{sec:design}.
Since obtaining the traffic distribution is not the focus of this work, we simply use the statistical information instead.
The \namea evolution and routing calculation are completed offline, with the resulting choices hard-coded into \nameb.
In our experiments, we focus on the performance of \nameb, which directly represents \name's effectiveness.
For comparison, we implement a series of typical routing schemes (as discussed in \S \ref{subsec:bg_routing}), covering deterministic (XY), oblivious (O1Turn, Valiant, ROMM), and adaptive (Odd-Even \cite{oddeven}) routing.

Throughput and latency are two basic metrics to judge the overall performance of routing algorithms.
We use the coefficient of variation (CV) of loads on different nodes to judge the severity of load imbalance (referred to as LCV).
The higher the LCV, the more severe the load imbalance.
Besides, we hypothesize a virtual ``reorder buffer'' at each destination to store out-of-order flits.
At a given moment, the maximum occupancy of all reorder buffers is referred to as the network's Reorder Value, which directly reflects the reordering overhead caused by out-of-order transmission.

\subsection{Synthetic Network Traffic} \label{subsec:synthetic}

\setlength{\tabcolsep}{3pt}
\begin{table}[t]
    \begin{tabular}{ccccccc}
        \hline
         & XY & O1Turn & Valiant & ROMM & \nameb\ \\\hline
        2DMesh+UN   & 0.29 & 0.28 & 0.35 & 0.46 & \textbf{0.20}\\\hline
        Edge I/O+UN & 0.28 & 0.36 & 0.33 & 0.19 & \textbf{0.08} \\\hline
        Edge I/O+OV & 0.36 & 0.63 & 0.37 & 0.30 & \textbf{0.17} \\\hline
    \end{tabular}
    \caption{LCVs of different routing algorithms.}
    \label{tab:revisit}
\end{table}

We first revisit the simple testcase in \S \ref{subsec:motiexp} to check the precision of \namea and the effectiveness of \nameb.
After that, typical traffic patterns are used to evaluate \nameb's overall performance.

\FBF{Revisiting the simple testcase.}A revisit to the simple testcase in \S \ref{subsec:motiexp} is conducted.
We investigate different nodes' $w_{NR}$ values and data forwarding rates (with \nameb\ adopted), and the results are included in Figure \ref{fig:load_distri}.
As the solid line indicates, \namea properly sets the $w_{NR}$s of each node, accurately estimating the overall trend in load distribution.
By making routing decisions `against' $w_{NR}$s, \nameb\ (the dashed curve) can smooth the imbalance of load distribution among different nodes.
According to our statistics (Table \ref{tab:revisit}), \nameb\ delivers significantly lower LCVs than other deterministic or oblivious routing algorithms, validating its effectiveness in load balancing.

\FBF{Performance under typical traffic patterns.}Four typical traffic patterns are used: Uniform, Shuffle, Permutation, and Overturn \cite{Dally2004principles}.
The results are shown in Figure \ref{fig:synthetic}.
\nameb's throughput significantly outperformed the deterministic and oblivious routing algorithms, and is similar to that of Odd-Even.
Under uniform traffic, the achieved throughput is 42.9\% higher than XY routing.
At the same time, it can be noticed that although Odd-Even performs well in benign cases like uniform and permutation, its performance degrades significantly under overturn traffic.
In this adverse case, nodes are universally overloaded, and it may be sensible to route according to the meticulously extracted load trend, instead of frequently adapting to local load information.

Besides overall performance, the impact of reordering is also evaluated (3rd row in Figure \ref{fig:synthetic}).
When the network load is low, data packets arrive sparsely and are rarely delivered out of order.
However, with the increase in network load, queue accumulation emerges, and packets arrive consecutively.
In this scenario, dynamic path switching can introduce significant out-of-order overhead, while the quasi-dynamic \nameb\ is free from this problem.

\subsection{Realistic Application Workload} \label{subsec:trace}
Besides synthetic network traffic, a realistic application workload is also evaluated.
Specifically, we trace the traffic between different ports of a switch, which operates as a leaf switch in an ns-3 simulated Clos network, and a realistic workload is applied to the Clos network \cite{li2019hpcc}.
The results are shown in Figure \ref{fig:switchtrace} (throughput is not compared because all workloads can be served within a finite time).

Due to the lack of real-time adaptiveness, the LCVs of \nameb\ are more dispersed than Odd-Even, but the mean value is lower than XY routing (by 14.5\%, from 0.568 to 0.486) and other routing algorithms except Valiant, indicating that \nameb\ can balance network load well most of the time.
From the perspective of network latency, \nameb\ performs significantly better than other deterministic and oblivious routing schemes, and even slightly better than Odd-Even.
Compared with XY routing, the average and maximum latencies are reduced by 86.4\% (from 149.6 to 20.4) and 95.3\% (from 572 to 27), respectively.
Besides, \nameb\ is free from out-of-order transmission, which saves significant overhead at runtime (in this test case, traffic load is generally light, but other dynamic routing schemes still experience non-negligible out-of-order).
Furthermore, \nameb\ is inherently free from out-of-order transmission, which yields substantial runtime overhead savings. Notably, although the traffic load is generally light in this test case, other dynamic routing schemes still incur out-of-order transmission to some extent.
\section{Conclusion}
In this paper, we propose \name, a quasi-dynamic routing scheme that addresses the long-standing dilemma between static and dynamic approaches. Rather than relying on dynamically collected load information, \name uses quasi-dynamically predicted load trends to guide routing decisions. It leverages \namea to extract long-term load information from the NoC topology and traffic distribution, and uses this to steer \nameb's routing choices. As a result, \name preserves the simplicity and predictability of static routing while effectively adapting to long-term load distribution to improve load balance.


\bibliographystyle{ACM-Reference-Format}
\bibliography{references}

\end{document}